\documentstyle[12pt]{art12}
\makeatletter
\let\reset@font\empty

\def\indexname{Index}
\def\figurename{Figure}
\def\tablename{Table}
\def\abstractname{Abstract}

\def\@ptsize{0}
\@namedef{ds@11pt}{\def\@ptsize{1}}
\@namedef{ds@12pt}{\def\@ptsize{2}}
\def\ds@twoside{\@twosidetrue
           \@mparswitchtrue}

\def\ds@draft{\overfullrule 5\p@}

\newif\if@titlepage \@titlepagefalse
\def\ds@titlepage{\@titlepagetrue}

\def\ds@twocolumn{\@twocolumntrue}

\newdimen\mathindent
\newif\ifletter
\newif\ifpmb
\newlength{\varind}
\newlength{\figdepth}
\newlength{\figwidth}
\newlength{\secfigwidth}

\newlength{\indentedwidth}

\newcounter{jnl}
\newcounter{yr}
\newcounter{tabtype}
\newcounter{figtype}
\newcounter{eqnval}
\@twosidetrue
\def\ds@draft{\overfullrule 5\p@}
\@options

\def\@normalsize{\@setsize\normalsize{16pt}\xiipt\@xiipt
  \abovedisplayskip 12pt plus3pt minus6pt
  \belowdisplayskip \abovedisplayskip
  \abovedisplayshortskip \z@ plus4pt
  \belowdisplayshortskip 7pt plus4pt minus4pt}
\def\small{\@setsize\small{14pt}\xipt\@xipt
  \abovedisplayskip 10pt plus 3pt minus 4pt
  \belowdisplayskip \abovedisplayskip
  \abovedisplayshortskip \z@ plus3pt
  \belowdisplayshortskip 5pt plus3pt minus 3pt
  \def\@listi{\topsep 5pt plus 3pt minus 3pt\parsep 0pt plus 1pt
         \itemsep \parsep}}
\def\footnotesize{\@setsize\footnotesize{14pt}\xpt\@xpt
  \abovedisplayskip 7pt plus 3pt minus 4pt
  \belowdisplayskip \abovedisplayskip
  \abovedisplayshortskip \z@ plus 2pt
  \belowdisplayshortskip 3pt plus 1pt minus2pt
  \def\@listi{\topsep 4pt plus 2pt minus 2pt\parsep 0pt plus 1pt
         \itemsep \parsep}}
\def\scriptsize{\@setsize\scriptsize{13pt}\ixpt\@ixpt}
\def\tiny{\@setsize\tiny{10pt}\viipt\@viipt}
\def\large{\@setsize\large{18pt}\xivpt\@xivpt}
\def\Large{\@setsize\Large{22pt}\xviipt\@xviipt}
\def\LARGE{\@setsize\LARGE{25pt}\xxpt\@xxpt}
\def\huge{\@setsize\huge{30pt}\xxvpt\@xxvpt}
\def\Huge{\@setsize\Huge{30pt}\xxvpt\@xxvpt}
\normalsize
\skip\footins 12\p@ plus 4\p@ minus 2\p@
\@beginparpenalty -\@lowpenalty
\@endparpenalty -\@lowpenalty
\@itempenalty -\@lowpenalty

\@noskipsecfalse   

\def\section{\@startsection{section}{1}{\z@}{-3.5ex plus -1ex minus
 -.2ex}{2.3ex plus .2ex}{\noindent\reset@font\normalsize\bf\raggedright}}
\def\subsection{\@startsection{subsection}{2}{\z@}{-3.25ex plus -1ex minus
 -.2ex}{1.5ex plus .2ex}{\noindent\reset@font
  \normalsize\it\raggedright\nohyphens}}
\def\subsubsection{\@startsection{subsubsection}{3}{\z@}{-3.25ex plus
-1ex minus -.2ex}{-1em}{\reset@font\normalsize\it\nohyphens}}
\def\paragraph{\@startsection
 {paragraph}{4}{\z@}{3.25ex plus 1ex minus
.2ex}{-1em}{\reset@font\normalsize\it}}
\def\subparagraph{\@startsection
 {subparagraph}{4}{\parindent}{3.25ex plus 1ex minus
 .2ex}{-1em}{\reset@font\normalsize\it}}

\def\@sect#1#2#3#4#5#6[#7]#8{\ifnum #2>\c@secnumdepth
     \let\@svsec\@empty\else
     \refstepcounter{#1}\edef\@svsec{\csname the#1\endcsname.\hskip 1em}\fi
     \@tempskipa #5\relax
      \ifdim \@tempskipa>\z@
        \begingroup #6\relax
          \noindent{\hskip #3\relax\@svsec}{\interlinepenalty \@M #8\par}%
        \endgroup
       \csname #1mark\endcsname{#7}\addcontentsline
         {toc}{#1}{\ifnum #2>\c@secnumdepth \else
                      \protect\numberline{\csname the#1\endcsname}\fi
                    #7}\else
        \def\@svsechd{#6\hskip #3\relax  
                   \@svsec #8\csname #1mark\endcsname
                      {#7}\addcontentsline
                           {toc}{#1}{\ifnum #2>\c@secnumdepth \else
                             \protect\numberline{\csname the#1\endcsname}\fi
                       #7}}\fi
     \@xsect{#5}}
\def\@ssect#1#2#3#4#5{\@tempskipa #3\relax
   \ifdim \@tempskipa>\z@
     \begingroup #4\noindent{\hskip #1}{\interlinepenalty
   \@M #5\par}\endgroup
   \else \def\@svsechd{#4\hskip #1\relax #5}\fi
    \@xsect{#3}}

\def\appendix{\@@par
 \setcounter{section}{0}
 \setcounter{subsection}{0}
 \setcounter{subsubsection}{0}
 \setcounter{equation}{0}
 \setcounter{figure}{0}
 \setcounter{table}{0}
 \def\thesection{Appendix \Alph{section}}
 \def\theequation{\ifnumbysec
      \Alph{section}.\arabic{equation}\else
      \Alph{section}\arabic{equation}\fi}
 \def\thetable{\ifnumbysec
      \Alph{section}\arabic{table}\else
      A\arabic{table}\fi}
 \def\thefigure{\ifnumbysec
      \Alph{section}\arabic{figure}\else
      A\arabic{figure}\fi}}

\leftmargin\leftmargini
\labelwidth\leftmargini\advance\labelwidth-\labelsep
\def\@listI{\leftmargin\leftmargini \parsep 4\p@ plus2\p@ minus\p@
\topsep 8\p@ plus2\p@ minus4\p@
\itemsep 4\p@ plus2\p@ minus\p@}

\let\@listi\@listI
\@listi

\def\@listii{\leftmargin\leftmarginii
 \labelwidth\leftmarginii\advance\labelwidth-\labelsep
 \topsep 3\p@ plus 1\p@ minus 1\p@
 \parsep 0\p@ plus 1\p@
 \itemsep \parsep}
\def\@listiii{\leftmargin\leftmarginiii
 \labelwidth\leftmarginiii\advance\labelwidth-\labelsep
 \topsep 2\p@ plus 1\p@ minus 1\p@
 \parsep \z@ \partopsep 1\p@ plus 0\p@ minus 1\p@
 \itemsep \topsep}
\def\@listiv{\leftmargin\leftmarginiv
 \labelwidth\leftmarginiv\advance\labelwidth-\labelsep}
\def\@listv{\leftmargin\leftmarginv
 \labelwidth\leftmarginv\advance\labelwidth-\labelsep}
\def\@listvi{\leftmargin\leftmarginvi
 \labelwidth\leftmarginvi\advance\labelwidth-\labelsep}

\def\hexnumber@#1{\ifcase#1 0\or 1\or 2\or 3\or 4\or 5\or 6\or 7\or 8\or
 9\or A\or B\or C\or D\or E\or F\fi}
\edef\bffam@{\hexnumber@\bffam}
\mathchardef\bGamma "0\bffam@00
\mathchardef\bDelta "0\bffam@01
\mathchardef\bTheta "0\bffam@02
\mathchardef\bLambda "0\bffam@03
\mathchardef\bXi "0\bffam@04
\mathchardef\bPi "0\bffam@05
\mathchardef\bSigma "0\bffam@06
\mathchardef\bUpsilon "0\bffam@07
\mathchardef\bPhi "0\bffam@08
\mathchardef\bPsi "0\bffam@09
\mathchardef\bOmega "0\bffam@0A

\def\theenumi{\roman{enumi}}

\def\theenumii{\alph{enumii}}
\def\p@enumii{\theenumi.}

\def\theenumiii{\arabic{enumiii}}
\def\p@enumiii{\p@enumii.\theenumii}

\def\p@enumiv{\p@enumiii.\theenumiii}

\def\labelitemi{$\m@th\bullet$}

\def\labelitemiii{$\m@th\ast$}
\def\labelitemiv{$\m@th\cdot$}

\def\verse{\let\\=\@centercr
 \list{}{\itemsep\z@ \itemindent -1.5em\listparindent \itemindent
 \rightmargin\leftmargin\advance\leftmargin 1.5em}\item[]}

\def\quotation{\list{}{\listparindent 1.5em
 \itemindent\listparindent
 \rightmargin\leftmargin\parsep 0\p@ plus 1\p@}\item[]}

\def\descriptionlabel#1{\hspace\labelsep \bf #1}
\def\description{\list{}{\labelwidth\z@ \itemindent-\leftmargin
 \let\makelabel\descriptionlabel}}

\def\enumerate{\ifnum \@enumdepth >3 \@toodeep\else
      \advance\@enumdepth \@ne
      \edef\@enumctr{enum\romannumeral\the\@enumdepth}\list
      {\csname label\@enumctr\endcsname}{\usecounter
        {\@enumctr}\def\makelabel##1{##1\hss}}\fi}
\def\itemize{\ifnum \@itemdepth >3 \@toodeep\else \advance\@itemdepth \@ne
\edef\@itemitem{labelitem\romannumeral\the\@itemdepth}%
\list{\csname\@itemitem\endcsname}{\def\makelabel##1{##1\hss}\topsep=3pt
  \parsep=0pt\listparindent=0pt\itemsep=0pt\partopsep=0pt\rightmargin=0pt
  }\fi}
\tabbingsep \labelsep
\skip\@mpfootins = \skip\footins
\def\titlepage{\@restonecolfalse\if@twocolumn\@restonecoltrue\onecolumn
     \else \newpage \fi \thispagestyle{myheadings}\c@page\z@}

\def\endtitlepage{\if@restonecol\twocolumn \else \newpage \fi}

\newcounter {section}
\newcounter {subsection}[section]
\newcounter {subsubsection}[subsection]
\newcounter {paragraph}[subsubsection]
\newcounter {subparagraph}[paragraph]

\def\thesection {\arabic{section}}

\def\@chapapp{Section}

\def\@pnumwidth{1.55em}
\def\@tocrmarg {2.55em}
\def\@dotsep{4.5}
\def\tableofcontents{\@restonecolfalse\if@twocolumn\@restonecoltrue
 \onecolumn\fi\section*{Contents}{}\thispagestyle{empty}
 \@starttoc{toc}\if@restonecol\twocolumn\fi}
\def\l@section{\@dottedtocline{1}{1.5em}{2.3em}}
\def\l@subsection{\@dottedtocline{2}{3.8em}{3.2em}}
\def\l@subsubsection{\@dottedtocline{3}{7.0em}{4.1em}}
\def\l@paragraph{\@dottedtocline{4}{10em}{5em}}
\def\l@subparagraph{\@dottedtocline{5}{12em}{6em}}
\def\listoffigures{\@restonecolfalse\if@twocolumn\@restonecoltrue\onecolumn
 \fi\section*{List of Figures\@mkboth
 {LIST OF FIGURES}{LIST OF FIGURES}}\@starttoc{lof}\if@restonecol\twocolumn
 \fi}
\def\l@figure{\@dottedtocline{1}{1.5em}{2.3em}}
\def\listoftables{\@restonecolfalse\if@twocolumn\@restonecoltrue\onecolumn
 \fi\section*{List of Tables\@mkboth
 {LIST OF TABLES}{LIST OF TABLES}}\@starttoc{lot}\if@restonecol\twocolumn
 \fi}
\let\l@table\l@figure
\def\@dottedtocline#1#2#3#4#5{\ifnum #1>\c@tocdepth \else
  \vskip \z@ plus .2\p@
  {\leftskip #2\relax \rightskip \@tocrmarg \parfillskip -\rightskip
    \parindent #2\relax\@afterindenttrue
   \interlinepenalty\@M
   \leavevmode
   \@tempdima #3\relax \advance\leftskip \@tempdima
   \hbox{}\hskip -\leftskip
    #4\nobreak\hfill \nobreak \hbox to\@pnumwidth{\hfil
   \rm #5}\@@par}\fi}

\@addtoreset{footnote}{page}
\long\def\@makefntext#1{\parindent 1em\noindent
 \makebox[1em][l]{\footnotesize\rm$\m@th{\fnsymbol{footnote}}$}%
 \footnotesize\rm #1}
\def\@makefnmark{\hbox{${\fnsymbol{footnote}}\m@th$}}
\def\@thefnmark{\fnsymbol{footnote}}
\def\footnote{\@ifnextchar[{\@xfootnote}{\stepcounter{\@mpfn}%
       \begingroup\let\protect\noexpand
       \xdef\@thefnmark{\thempfn}\endgroup
     \@footnotemark\@footnotetext}}
\def\@fnsymbol#1{\ifcase#1\or \dagger\or \ddagger\or \S\or
   \|\or \P\or ^{+}\or ^{\tsty *}\or \sharp
   \or \dagger\dagger \else\@ctrerr\fi\relax}

\def\[{\relax\ifmmode\@badmath\else
 \begin{trivlist}
 \@beginparpenalty\predisplaypenalty
 \@endparpenalty\postdisplaypenalty
 \item[]\leavevmode
 \hbox to\linewidth\bgroup$ \displaystyle
 \hskip\mathindent\bgroup\fi}
\def\]{\relax\ifmmode \egroup $\hfil \egroup \end{trivlist}\else \@badmath \fi}
\def\equation{\@beginparpenalty\predisplaypenalty
 \@endparpenalty\postdisplaypenalty
\refstepcounter{equation}\trivlist \item[]\leavevmode
 \hbox to\linewidth\bgroup $ \displaystyle
\hskip\mathindent}
\def\endequation{$\hfil \displaywidth\linewidth\@eqnnum\egroup \endtrivlist}
\def\eqnarray{\stepcounter{equation}\let\@currentlabel=\theequation
\global\@eqnswtrue
\global\@eqcnt\z@\tabskip\mathindent\let\\=\@eqncr
\abovedisplayskip\topsep\ifvmode\advance\abovedisplayskip\partopsep\fi
\belowdisplayskip\abovedisplayskip
\belowdisplayshortskip\abovedisplayskip
\abovedisplayshortskip\abovedisplayskip
$$\halign to
\linewidth\bgroup\@eqnsel$\displaystyle\tabskip\z@
 {##{}}$&\global\@eqcnt\@ne $\displaystyle{{}##{}}$\hfil    
 &\global\@eqcnt\tw@ $\displaystyle{{}##}$\hfil
 \tabskip\@centering&\llap{##}\tabskip\z@\cr}
\def\endeqnarray{\@@eqncr\egroup
 \global\advance\c@equation\m@ne$$\global\@ignoretrue }
\newcommand{\jl}[1]{\setcounter{jnl}{#1}%
    \ifnum\thejnl=12\global\pmbtrue\fi
    \ifnum\thejnl=15\global\pmbtrue\fi}
\def\journal{\ifnum\thejnl=1 J. Phys.\ A: Math.\ Gen.\
        \else\ifnum\thejnl=2 J. Phys.\ B: At.\ Mol.\ Opt.\ Phys.\
        \else\ifnum\thejnl=3 J. Phys.:\ Condens. Matter\
        \else\ifnum\thejnl=4 J. Phys.\ G: Nucl.\ Part.\ Phys.\
        \else\ifnum\thejnl=5 Inverse Problems\
        \else\ifnum\thejnl=6 Class. Quantum Grav.\
        \else\ifnum\thejnl=7 Network\
        \else\ifnum\thejnl=8 Nonlinearity\
        \else\ifnum\thejnl=9 Quantum Opt.\
        \else\ifnum\thejnl=10 Waves in Random Media\
        \else\ifnum\thejnl=11 Pure Appl. Opt.\
        \else\ifnum\thejnl=12 Phys. Med. Biol.\ %
        \else\ifnum\thejnl=13 Modelling Simul.\ Mater.\ Sci.\ Eng.\
        \else\ifnum\thejnl=14 Plasma Phys. Control. Fusion\
        \else\ifnum\thejnl=15 Physiol. Meas.\
        \else\ifnum\thejnl=16 Sov.\ Lightwave Commun.\
        \else\ifnum\thejnl=17 High Perform.\ Polym.\
        \else\ifnum\thejnl=18 J.\ Hard Mater.\
        \else\ifnum\thejnl=19 J.\ Phys.\ D: Appl.\ Phys.\
        \else\ifnum\thejnl=20 Supercond.\ Sci.\ Technol.\
        \else\ifnum\thejnl=21 Semicond.\ Sci.\ Technol.\
        \else\ifnum\thejnl=22 Nanotechnology\
        \else\ifnum\thejnl=23 Meas.\ Sci.\ Technol.\
        \else\ifnum\thejnl=24 Plasma Source Sci.\ Technol.\
        \else\ifnum\thejnl=25 Smart Mater.\ Struct.\
        \else\ifnum\thejnl=26 J.\ Micromech.\ Microeng.\
        \else\ifnum\thejnl=27 Distrib.\ Syst.\ Engng\
\else Institute of Physics Publishing
\fi\fi\fi\fi\fi\fi\fi\fi\fi\fi\fi\fi\fi\fi\fi
\fi\fi\fi\fi\fi\fi\fi\fi\fi\fi\fi\fi}

\def\catchline{\hfill}

\def\cpyrtline{\hfill}
\def\maketitle{\vspace*{\baselineskip}\vspace{0\p@ plus1fil}
    \noindent Short title: \@shorttitle\par
    \@submitted
    \vspace*{\baselineskip}
    \noindent\today\par\newpage}
\def\@rticle#1#2{\thispagestyle{myheadings}%
     \vspace*{.5pc}%
    {\parindent=\mathindent \bf #1\par}%
     \vspace*{1.5pc}%
    {\exhyphenpenalty=10000\hyphenpenalty=10000
     \Large\raggedright\noindent
     \bf#2\par}\def\@shorttitle{#1}\futurelet\next\sh@rttitle}%
\def\title#1{\def\@shorttitle{#1}%
    \thispagestyle{myheadings}%
    \vspace*{3pc}{\exhyphenpenalty=10000\hyphenpenalty=10000
    \Large\raggedright\noindent
    \bf#1\par}\futurelet\next\sh@rttitle}

\def\article#1#2{\@rticle{#1}{#2}}
\def\review#1{\@rticle{REVIEW \ifpmb\else ARTICLE\fi}{#1}}
\def\topical#1{\@rticle{TOPICAL REVIEW}{#1}}
\def\ireview#1{\@rticle{INTRODUCTORY REVIEW}{#1}}
\def\comment#1{\@rticle{COMMENT}{#1}}
\def\note#1{\@rticle{NOTE}{#1}}
\def\prelim#1{\@rticle{PRELIMINARY COMMUNICATION}{#1}}
\def\letter#1{\@rticle{LETTER TO THE EDITOR}{#1}}
\def\sh@rttitle{\ifx\next[\let\next=\sh@rt
                \else\let\next=\f@ll\fi\next}
\def\sh@rt[#1]{\gdef\@shorttitle{#1}}
\def\f@ll{}
\renewcommand{\author}[1]{\vspace*{1.5pc}%
   \begin{indented}%
   \item[]\normalsize\ifnum\thejnl=8\bf\else\rm\fi\raggedright#1
   \end{indented}%
   \smallskip}
\def\abstract{\vspace{16pt plus3pt minus3pt}
   \begin{indented}
   \item[]{\bf \abstractname.}\quad\rm\ignorespaces}%
\def\endabstract{\end{indented}\vspace{18\p@ plus18\p@}}
\def\submitted{\def\@submitted{\vspace{\baselineskip}%
     \noindent Submitted to: \journal\par}}
\def\@submitted{}
\def\nosections{\vspace{30\p@ plus12\p@ minus12\p@}
    \noindent\ignorespaces}
\def\ack{\ifletter\bigskip\noindent\ignorespaces\else
    \section*{Acknowledgments}\fi}

\newif\ifnumbysec
\def\theequation{\ifnumbysec
      \arabic{section}.\arabic{equation}\else
      \arabic{equation}\fi}
\def\eqnobysec{\numbysectrue\@addtoreset{equation}{section}}
\newenvironment{ceqno}{\begin{ceqno}}{\end{ceqno}}
\def\ceqno{\begin{equation}\begin{array}{@{}*{4}{l}}\dsty}
\def\endceqno{\end{array}\end{equation}}
\def\eqalign#1{\null\vcenter{\def\\{\cr}\openup\jot\m@th
  \ialign{\strut$\displaystyle{##}$\hfil&$\displaystyle{{}##}$\hfil
      \crcr#1\crcr}}\,}
\def\eqalignno#1{\displ@y \tabskip\z@skip
  \halign to\displaywidth{\hspace{5pc}$\@lign\displaystyle{##}$%
    \tabskip\z@skip
    &$\@lign\displaystyle{{}##}$\hfill\tabskip\@centering
    &\llap{$\@lign\hbox{\rm##}$}\tabskip\z@skip\crcr
    #1\crcr}}
\def\cases#1{%
     \left\{\,\vcenter{\def\\{\cr}\normalbaselines\openup1\jot\m@th%
     \ialign{\strut$\displaystyle{##}\hfil$&\tqs
     \rm##\hfil\crcr#1\crcr}}\right.}%
\def\tabular{\def\@halignto{}\@tabular}
\newcommand{\Table}[1]{\def\t@blecap{\caption{#1}}%
   \setcounter{tabtype}{1}\futurelet\next\t@bplace}
\newcommand{\widetable}[1]{\def\t@blecap{\caption{#1}}%
   \setcounter{tabtype}{2}\futurelet\next\t@bplace}
\newcommand{\fulltable}[1]{\def\t@blecap{\caption{#1}}%
   \setcounter{tabtype}{3}\futurelet\next\t@bplace}%
\def\t@bplace{\ifx\next[\let\next=\@tabpl
                 \else\let\next=\@tabnopl\fi\next}
\def\@tabpl[#1]{\begin{table}[#1]\@t@bsize}
\def\@tabnopl{\begin{table}\@t@bsize}
\def\@t@bsize{\ifnum\thetabtype=3\begin{varindent}{0pt}%
   \else\begin{varindent}{\mathindent}\fi
   \t@blecap\lineup\item[]
   \ifnum\thetabtype=1
        \begin{tabular}{@{}l*{15}{l}}
   \else\ifnum\thetabtype=2
        \begin{tabular*}{\indentedwidth}{@{}l*{15}{@{\extracolsep{0pt
plus12pt}}l}}
   \else\begin{tabular*}{\textwidth}{@{}l*{15}{@{\extracolsep{0pt plus12pt}}l}}
   \fi\fi}
\def\endtab{\ifnum\thetabtype=1\end{tabular}
   \else\end{tabular*}\fi\end{varindent}\end{table}}

\def\lineup{\def\0{\hbox{\phantom{\footnotesize\rm 0}}}%
    \def\m{\hbox{$\phantom{-}$}}%
    \def\-{\llap{$-$}}}
\long\def\@makecaption#1#2{\vskip 10\p@
 \ifnum\thefigtype=2\begin{varindent}{\@figindent}
 \item[]{\bf #1.} #2
 \end{varindent}\else
 \ifnum\thefigtype=3
 \footnotesize\rm{\bf #1.} #2\else
 \begin{indented}
 \item[]{\bf #1.} #2
 \end{indented}\fi\fi}
\newcommand{\Figure}[1]{\setcounter{figtype}{1}%
    \def\figspace{}\def\figcap{\caption{#1}}%
    \futurelet\next\@figplace}
\def\@figplace{\ifx\next[\let\next=\@figpl
                 \else\let\next=\@fignopl\fi\next}
\def\@figpl[#1]{\begin{figure}[#1]
   \figspace
   \figcap
   \end{figure}}
\def\@fignopl{\begin{figure}
   \figspace
   \figcap
   \end{figure}}

\newcommand{\sidecap}[3]{\setcounter{figtype}{2}%
    \setlength{\figdepth}{#1}\def\@figindent{#2}%
    \def\sidedc@p{\caption{#3}}%
    \futurelet\next\@sidecapplace}
\def\@sidecapplace{\ifx\next[\let\next=\@sidecappl
                 \else\let\next=\@sidecapnopl\fi\next}
\def\@sidecappl[#1]{\begin{figure}[#1]
    \vbox to\figdepth{\vfill
    \sidedc@p}%
    \setcounter{figtype}{1}\end{figure}}
\def\@sidecapnopl{\begin{figure}
    \vbox to\figdepth{\vfill
    \sidedc@p}%
    \setcounter{figtype}{1}\end{figure}}
\newcommand{\side}[3]{\setcounter{figtype}{3}%
    \setlength{\figdepth}{#1}\setlength{\figwidth}{15pc}
    \setlength{\secfigwidth}{15pc}
    \def\firstc@p{\caption{#2}}\def\secondc@p{\caption{#3}}
    \futurelet\next\@sideplace}
\def\@sideplace{\ifx\next[\let\next=\@sidepl
                 \else\let\next=\@sidenopl\fi\next}
\def\@sidepl[#1]{\begin{figure}[#1]
    \vspace*{1.5pc}\vspace*{\figdepth}
    \parbox[t]{\figwidth}{\firstc@p}\hspace*{1pc}%
    \parbox[t]{\secfigwidth}{\secondc@p}
    \setcounter{figtype}{1}\end{figure}}
\def\@sidenopl{\begin{figure}
    \vspace*{1.5pc}\vspace*{\figdepth}
    \parbox[t]{\figwidth}{\firstc@p}\hspace*{1pc}%
    \parbox[t]{\secfigwidth}{\secondc@p}
    \setcounter{figtype}{1}\end{figure}}
\newcommand{\varside}[4]{\setcounter{figtype}{3}%
    \setlength{\figdepth}{#1}\setlength{\figwidth}{#2}%
    \setlength{\secfigwidth}{30pc}
    \addtolength{\secfigwidth}{-\figwidth}
    \def\firstc@p{\caption{#3}}\def\secondc@p{\caption{#4}}
    \futurelet\next\@sideplace}

\newcounter{figure}
\def\thefigure{\@arabic\c@figure}
\def\fps@figure{htbp}
\def\ftype@figure{1}
\def\ext@figure{lof}
\def\fnum@figure{\figurename~\thefigure}
\def\figure{\@float{figure}}
\let\endfigure\end@float
\@namedef{figure*}{\@dblfloat{figure}}
\@namedef{endfigure*}{\end@dblfloat}
\newcounter{table}
\def\thetable{\@arabic\c@table}
\def\fps@table{htbp}
\def\ftype@table{2}
\def\ext@table{lot}
\def\fnum@table{\tablename~\thetable}
\def\table{\@float{table}}
\let\endtable\end@float
\@namedef{table*}{\@dblfloat{table}}
\@namedef{endtable*}{\end@dblfloat}

\def\thebibliography#1{\list
 {\hfil[\arabic{enumi}]}{\topsep=0\p@\parsep=0\p@
 \partopsep=0\p@\itemsep=0\p@
 \labelsep=5\p@\itemindent=-10\p@
 \settowidth\labelwidth{\footnotesize[#1]}%
 \leftmargin\labelwidth
 \advance\leftmargin\labelsep
 \advance\leftmargin -\itemindent
 \usecounter{enumi}}\footnotesize
 \def\newblock{\ }
 \sloppy\clubpenalty4000\widowpenalty4000
 \sfcode`\.=1000\relax}

\def\numrefs#1{\begin{thebibliography}{#1}}
\def\endnumrefs{\end{thebibliography}}

\def\thereferences{\list{}{\topsep=0\p@\parsep=0\p@
 \partopsep=0\p@\itemsep=0\p@\labelsep=0\p@\itemindent=-18\p@
\labelwidth=0\p@\leftmargin=18\p@
}\footnotesize\rm
\def\newblock{\ }
\sloppy\clubpenalty4000\widowpenalty4000
\sfcode`\.=1000\relax
}
\newenvironment{harvard}{\list{}{\topsep=0\p@\parsep=0\p@
\partopsep=0\p@\itemsep=0\p@\labelsep=0\p@\itemindent=-18\p@
\labelwidth=0\p@\leftmargin=18\p@
}\footnotesize\rm
\def\newblock{\ }
\sloppy\clubpenalty4000\widowpenalty4000
\sfcode`\.=1000\relax}{\endlist}
\def\refs{\begin{harvard}}
\def\endrefs{\end{harvard}}
\newenvironment{indented}{\begin{indented}}{\end{indented}}
\newenvironment{varindent}[1]{\begin{varindent}{#1}}{\end{varindent}}
\def\indented{\list{}{\itemsep=0\p@\labelsep=0\p@\itemindent=0\p@
   \labelwidth=0\p@\leftmargin=\mathindent\topsep=0\p@\partopsep=0\p@
   \parsep=0\p@\listparindent=15\p@}\footnotesize\rm}

\def\varindent#1{\setlength{\varind}{#1}%
   \list{}{\itemsep=0\p@\labelsep=0\p@\itemindent=0\p@
   \labelwidth=0\p@\leftmargin=\varind\topsep=0\p@\partopsep=0\p@
   \parsep=0\p@\listparindent=15\p@}\footnotesize\rm}

\def\tabnotes{\ifnum\thetabtype=1\end{tabular}\else\end{tabular*}\fi}
\def\endtabnotes{\end{varindent}\end{table}}

\newif\if@restonecol
\def\theindex{\@restonecoltrue\if@twocolumn\@restonecolfalse\fi
\columnseprule \z@
\columnsep 35\p@\twocolumn[\section*{\indexname}]%
    \@mkboth{{\indexname}}{{\indexname}}%
    \parindent\z@
    \parskip\z@ plus.3\p@\relax\let\item\@idxitem}
\def\@idxitem{\par\hangindent 30\p@}
\def\subitem{\par\hangindent 30\p@ \hspace*{10\p@}}
\def\subsubitem{\par\hangindent 30\p@ \hspace*{20\p@}}
\def\endtheindex{\if@restonecol\onecolumn\else\clearpage\fi}
\def\indexspace{\par \vskip 10\p@ plus 5\p@ minus 3\p@\relax}

\mark{{}{}}

\def\ps@headings{\let\@mkboth\markboth
 \def\@oddfoot{}%
 \def\@evenfoot{}%
 \def\@evenhead{\makebox[\mathindent][l]{\normalsize\rm \thepage}%
  \normalsize\it\rightmark\hfill}%
 \def\@oddhead{\makebox[\mathindent][r]{\hfill}{\normalsize\it\leftmark}\hfill
  \normalsize\rm\thepage}%
}%

\def\ps@myheadings{\let\@mkboth\markboth
 \def\@oddhead{\catchline}%
 \def\@oddfoot{\cpyrtline}%
 \def\@evenhead{}%
 \def\@evenfoot{}%
}

\def\today{\ifcase\month\or
 January\or February\or March\or April\or May\or June\or
 July\or August\or September\or October\or November\or December\fi
 \space\number\day, \number\year}

\def\@begintheorem#1#2{\rm \trivlist \item[\hskip \labelsep{\it #1\ #2.}]}
\def\@opargbegintheorem#1#2#3{\rm \trivlist
      \item[\hskip \labelsep{\it #1\ #2\ (#3).}]}

\def\p@LaTeX{{L\kern-.3em\lower.1em\hbox{$^{\rm A}$}\kern-.15em%
    T\kern-.1667em\lower.7ex\hbox{E}\kern-.125emX}}
\newcommand{\text}[1]{\mbox{#1}}

\newcommand{\nohyphens}{\hyphenpenalty=10000\exhyphenpenalty=10000}

\renewcommand{\qquad}{\hspace*{25pt}}
\newcommand{\tqs}{\hspace*{25pt}}
\newcommand{\fl}{\hspace*{-\mathindent}}

\newcommand{\tr}{\mathop{\mathrm tr}\nolimits}

\def\pt(#1){({\it #1\/})}
\newcommand{\dsty}{\displaystyle}
\newcommand{\tsty}{\textstyle}

\def\;{\protect\psemicolon}
\def\psemicolon{\relax\ifmmode\mskip\thickmuskip\else\kern .3333em\fi}

\newcommand{\opencirc}{\raisebox{2\p@}{\;\circle{5}}}

\newcommand{\fullcirc}{\raisebox{-2\p@}{\Large$\bullet$}}

\newcommand{\boldarrayrulewidth}{1\p@}

\def\bhline{\noalign{\ifnum0=`}\fi\hrule \@height
\boldarrayrulewidth \futurelet \@tempa\@xhline}

\def\@xhline{\ifx\@tempa\hline\vskip \doublerulesep\fi
      \ifnum0=`{\fi}}

\newcommand{\ms}{\noalign{\vspace{3\p@ plus2\p@ minus1\p@}}}
\newcommand{\bs}{\noalign{\vspace{6\p@ plus2\p@ minus2\p@}}}
\newcommand{\ns}{\noalign{\vspace{-3\p@ plus-1\p@ minus-1\p@}}}
\newcommand{\es}{\noalign{\vspace{6\p@ plus2\p@ minus2\p@}}\displaystyle}
\newcommand{\JPA}{{\em J. Phys. A: Math. Gen.} }

\ps@headings  \onecolumn
\makeatother
\begin{document}
\jl{1}
\begin{flushright}
SISSA Ref. 36/97/EP
\end{flushright}
\vspace*{1.0cm}
\begin{center}
   {\bf \Large Stochastic Models on a Ring  and Quadratic Algebras. \\
The Three Species Diffusion Problem.}
   \\[20mm]
Peter F. Arndt$\mbox{}^\star$,
Thomas Heinzel$\mbox{}^\star$
and Vladimir Rittenberg$\mbox{}^\diamond$
\\[7mm]
{SISSA, Via Beirut 2--4, 34014 Trieste, Italy\\[3mm]
permanent address:
Physikalisches Institut, Nu{\ss}allee 12,
53115 Bonn, Germany}
\\[2.2cm]
\end{center}
\renewcommand{\thefootnote}{\arabic{footnote}}
\addtocounter{footnote}{-1}
\vspace*{2mm}
%
The stationary state of a stochastic process on a ring can be
expressed using traces of monomials of an associative algebra
defined by quadratic relations.
If one considers only exclusion
processes one can restrict the type of algebras and
obtain recurrence relations for the traces.
This is possible only
if the rates satisfy certain compatibility conditions.
These
conditions are derived and the recurrence relations solved
giving representations of the algebras.
\vspace*{4.0cm}
\begin{flushleft}
cond-mat/9703182
\\
March 1997\\[1cm]
$\mbox{}^\star$ work supported by the DAAD programme HSP II--AUFE\\
$\mbox{}^\diamond$  work done with partial support of the EC TMR programme,
grant FMRX-CT96-0012
\end{flushleft}
\thispagestyle{empty}
\mbox{}
\newpage
\setcounter{page}{1}
\eqnobysec
\section{Introduction}
In the preceding paper \cite{AlDaRi} we
have considered the application of
quadratic algebras to stochastic problems with closed or open
boundaries, here we
study the case of periodic boundary conditions.
Again, we are going to be
interested in the (unnormalized)
probability distributions describing stationary states.
In the language of quantum chains, we are looking for ground states which
have momentum and energy zero.
A lot of work was already done looking for
matrix-product states in the case of periodic chains
\cite{HaNa,FaNaWe,NiKlZi}; the present
approach is different in the sense that it is based on the existence
of recurrence
relations which can be solved using representations of some quadratic
algebras.
The idea of this approach is not new \cite{DeJaLeSp,Ma} and all we did is to
pursue it in a consistent way in the case of models with three states.
As
a result one finds more solutions than were known before which can be used
either to repeat the
previous applications \cite{DeJaLeSp,Ma}
in a more general framework or to look for novel
applications.
The same approach can be extended to problems with more states.
Such an extension is not trivial.

{}From a mathematical point of view one has to
solve a well stated problem: given a certain class of quadratic algebras
one has to find those which are compatible with the trace operation.
One lesson one learns from the present work
is that, unexpectedly, in order to solve the periodic case one makes use
of Fock
representations, derived in the previous paper \cite{AlDaRi},
which were necessary
to solve the problem with closed or open boundaries.

We first consider the general case of $N$ species on a ring with $L$ sites and
use the notations of Ref.\cite{AlDaRi}.
On each site we take a stochastic variable
$\beta_k$ ($\beta=0,1,\dots ,N-1$ and $k=1,2,\dots ,L$), on each link $k$
between the sites $k$
and $k+1$ the rates $\Gamma^{\gamma_k\gamma_{k+1}}_{\beta_k\beta_{k+1}}$
give the probability per unit time for the
transition
\begin{equation}
\{\dots,\gamma_k,\gamma_{k+1},\dots\}\mapsto
            \{\dots,\beta_k,\beta_{k+1},\dots\}\; .
\end{equation}
The Hamiltonian associated with the master
equation is \cite{AlDaRi}:
\begin{equation}
\label{eqi}
H=-\sum^L_{k=1}
 \Gamma^{\alpha\beta}_{\gamma\delta}\,
     E^{\gamma\alpha}_k\, E^{\delta\beta}_{k+1}
\end{equation}
where the matrices $E_k$ act on the $k$th site and have matrix elements
%
%
\begin{equation}
\label{eqii}
\left(E^{\alpha\beta}\right)_{\gamma\delta}=\delta_{\alpha\gamma}
\delta_{\beta\delta}\;\; ,
\end{equation}
and the diagonal elements
$\Gamma^{\alpha\beta}_{\alpha\beta}$
are given by
\begin{equation}
\sum_{(\gamma,\delta)} \Gamma^{\alpha\beta}_{\gamma\delta}=0 \;\; .
\end{equation}
The site $L+1$ is identified with the first site.

Now it is trivial to show
that if we take $N$ matrices $D_\alpha$ ($\alpha=0,1,\dots ,N-1$)
and $N$ matrices $X_\alpha$
satisfying the quadratic algebra
\begin{equation}
\label{eqiii}
\sum_{\alpha ,\beta=0}^{N-1}
 \Gamma^{\alpha\beta}_{\gamma\delta}\,
 D_\alpha \, D_\beta=X_\gamma\, D_\delta- D_\gamma\, X_\delta
\qquad
(\gamma ,\delta=0,1,\dots ,N-1)
\end{equation}
then
\def\tr{{\rm Tr}}
\begin{equation}
\label{eqiv}
P_s=\tr \left(\prod^L_{k=1}\left(
\sum_{\alpha=0}^{N-1} D_\alpha u_\alpha^{(k)}
\right)\right)
\end{equation}
is a stationary state:
\begin{equation}
\label{eqv}
H\cdot P_s=0
\end{equation}
We have denoted by $u_\alpha^{(k)}$
($ \alpha=0,1,\dots ,N-1$ and $k=1,2,\dots ,L$)
the basis vectors in the
vector space associated to the $k$th site on which the basis matrices
$E^{\alpha,\beta}_k$
of Eq.(\ref{eqi}) act.
The trace operation in Eq.(\ref{eqiv}) is taken in the auxiliary space of the
$D_\alpha$ and $X_\alpha$ matrices.
We notice that the bulk algebra (\ref{eqiii})
is identical to the one encountered in
the previous paper.
The new thing is the appearance of the trace operation
in Eq.(\ref{eqiv}).

As opposed to the problem with closed and open boundaries where
the bulk algebra
was completed by the condition of the existence of a Fock representation
defined by the boundary conditions and where it was shown that $D$ and $X$
matrices can be derived once the bulk and boundary
rates are given \cite{KrSa}, in the present case
very little is known,
except that the bulk algebra exists since a representation for
the $D$'s and $X$'s is known \cite{KrSa}.
This representation however is pathologic in that the traces
of any monomial of $D$'s vanish.
It is also not clear if all stationary states can be obtained
through the ansatz (\ref{eqiii}) \cite{Kr}.

The remarkable thing about the algebras (\ref{eqiii}) is that,
if the ground state (\ref{eqiv}) is unique, all the
traces of monomials of degree $L$ and containing only $D_\alpha$'s
and no $X_\alpha$'s
are, up to a common factor, independent of
the representation of the algebra which implies that in order to
compute them one can take the smallest one.

Last but not least, let us observe that the cases
(see for example \cite{NiKlZi})
where the ansatz was applied correspond to representations
with $X_\alpha =0$ \cite{Kr}.
This leads us to polynomial algebras like in Sec.3 of Ref.\cite{AlDaRi}.

We will now restrict our problem looking only to simple exclusion processes.
This means that only the rates
$\Gamma^{\alpha\beta}_{\beta\alpha}=g_{\alpha\beta}$ and the diagonal
ones are non-zero.
Also, we will look
for solutions in which the $X_\alpha$ matrices are c-numbers $x_\alpha$.
This last
assumption will imply conditions on the rates $g_{\alpha\beta}$  .
The quadratic algebras
have now a simple form:
\begin{equation}
\label{eqvi}
\fl
g_{\alpha\beta}\, D_\alpha \, D_\beta-g_{\beta\alpha}\, D_\beta\,  D_\alpha =
x_\beta \, D_\alpha-x_\alpha\,  D_\beta
\qquad
(\alpha,\beta=0,1,\dots ,N-1)
\end{equation}
There are $N(N-1)/2$ relations with $N$ parameters $x_\alpha$ and $N$
generators $D_\alpha$.

The appearance of $N$ arbitrary parameters in the algebra can be
understood in the following way. The problem has a $U(1)^{N-1}$ symmetry
corresponding to the conservation of the number of particles of $N-1$
species (the remaining species are the vacancies).
The ground state is highly degenerate.
If one has a ring with $L$ sites, $P_s$ given by Eq.(\ref{eqiv})
has the following formal expression:
\begin{equation}
\label{eqai}
P_s=\sum_{n_\alpha} {d_0}^{n_0} {d_1}^{n_1} \cdots {d_{N-1}}^{n_{N-1}}\,
	A_{n_0,n_1,\ldots,n_{N-1}}
\end{equation}
where
\begin{equation}
\sum^{N-1}_{\alpha=0} n_{\alpha} = L
\end{equation}
The $d_{\alpha}$ are arbitrary parameters and
$A_{n_0,n_1,\ldots,n_{N-1}}$ are vectors on which $H$ acts.
In this way, in each sector,
given by the number $n_0$
of vacancies and $n_i$ of particles of type $i$,
the ground state can be identified.
That $P_s$ is indeed of the form (\ref{eqai}) can be seen from the invariance
of the algebra (\ref{eqvi}) under the transformation:
\begin{equation}
\label{eqaiii}
D_\alpha \mapsto d_\alpha \, D_\alpha \, , \qquad
x_\alpha \mapsto d_\alpha \, x_\alpha
\end{equation}
Let us stress again that in order to obtain the expression (\ref{eqai})
one can take any representation of the algebra.
We now turn our attention to the algebra (\ref{eqvi}).

What
we need is not only to have associative algebras but that the trace
operation exists.
Since the left hand side of the Eqs.(\ref{eqvi}) is quadratic and
the right hand side linear in the generators $D_\alpha$,
this implies recurrence
relations among traces of monomials and hence compatibility relations among
the rates $g_{\alpha\beta}$.
In order to solve this problem, our strategy was the
following: we have first considered monomials of a low degree (up to
five), got the compatibility relations and made sure that no new relations
occur from higher degree monomials by finding a representation with finite
traces for the algebra.
Once we have the algebra, for the physical
applications one can do calculations using either directly the algebra
and do formal manipulations under the trace operation or use the
representation.
This procedure is going to be explained in
detail in section 3 for the three-state case.
In the short section 2 we ask a simple question:
what are the conditions on the $g_{\alpha\beta}$
for arbitrary $N$ such that one has
one-dimensional representations? (one-dimensional representations
obviously have
a trace.) Finally, in section 4 we present our conclusions.

\section{One-dimensional representations}
The simplest examples of the algebras defined by Eq.(\ref{eqvi})
which have a trace
are those in which one has one-dimensional representations.
In order to
find them, we take $D_\alpha =d_\alpha$, arbitrary non-zero c-numbers.
It is convenient to introduce the
notations:
\begin{equation}
\label{eqvii}
a_{\alpha \beta}=g_{\alpha \beta}-g_{\beta \alpha}
\end{equation}
Using Eq.(\ref{eqvi}) one obtains:
\begin{equation}
\label{eqviii}
\frac{x_\alpha}{d_\alpha} - \frac{x_\beta}{d_\beta} = a_{\beta\alpha}
\qquad
(\alpha, \beta=0,1,\dots , N-1)
\end{equation}
These equations determine the parameters $x_\alpha$ once the $d_\alpha$ are
chosen.
One
obtains $(N-1)(N-2)/2$ conditions on the rates:
\begin{equation}
\label{eqix}
a_{0 \alpha}-a_{0\beta}=a_{\beta\alpha}
\qquad
(\alpha, \beta=1, 2,\dots , N-1)
\end{equation}
and the parameters $x_\alpha$ are
\begin{equation}
\label{eqx}
x_\alpha=d_\alpha (a_{0 \alpha}+x_0/ d_0) \qquad
(\alpha=1, 2,\dots , N-1)
\end{equation}
Notice that one of them can be chosen at will.
In Eqs.(\ref{eqix}) and (\ref{eqx}) we have singled out $\alpha=0$
as a matter of notational convenience.

The wave function, see
Eq.(\ref{eqiv}), is symmetric.
This observation is interesting for the following
reason.
In the case of simple exclusion processes, the Hamiltonian given
by Eq.(\ref{eqi}) has a $U(1)^{N-1}$ symmetry.
As already discussed, this corresponds to the conservation
of the $N-1$ types of particles hopping among vacancies.
The symmetric wave function, however,
corresponds to an $SU(N)$ representation given by the Young
tableau with one row and $L$ boxes ($L$ is the number of sites), although the
Hamiltonian does not have this symmetry.
{}From Eq.(\ref{eqix}) we also
see that for $N=2$ one has one-dimensional representations for any rates.

\section{The three-state algebras}
We will now study in detail the algebras given by Eq.(\ref{eqvi}) for
$N=3$. In
order to find them, we will consider several cases.
\subsection{$x_0$, $x_1$ and $x_2$ different from zero}
We define
\begin{equation}
\label{eti}
D_i=x_i\, E_i\qquad (i=1,2,3)
\end{equation}
and get the algebra:
\begin{eqnarray}
\label{etii}
g_{01}\, E_0\, E_1 - g_{10}\, E_1\, E_0 = E_0 - E_1 \nonumber\\
g_{20}\, E_2\, E_0 - g_{02}\, E_0\, E_2 = E_2 - E_0 \\
g_{12}\, E_1\, E_2 - g_{21}\, E_2\, E_1 = E_1 - E_2 \nonumber
\end{eqnarray}
{}From writing the recurrence relations for monomials of degree two and
three, the equations giving $\tr (E_0\,E_1\,E_2)$ and
$\tr (E_2\,E_1\,E_0)$ are
consistent only if
\begin{equation}
\label{etiii}
a_{01}-a_{02}=a_{21}
\end{equation}
or if
\begin{eqnarray}
a_{12}\, (g_{10}\, g_{20} - g_{01}\, g_{02})\, \tr (E_0) + \nonumber\\
a_{20}\, (g_{01}\, g_{21} - g_{10}\, g_{12})\, \tr (E_1) + \nonumber\\
a_{01}\, (g_{02}\, g_{12} - g_{20}\, g_{21})\, \tr (E_2) = 0\label{etiv}
\end{eqnarray}
Eq.(\ref{etiii}) gives the condition to have a one-dimensional representation,
see (\ref{eqix}).
Eq.(\ref{etiv}) however is new.
We use (\ref{etiv}) to express
$\tr (E_0)$ in terms of $\tr (E_1)$ and $\tr (E_2)$
and look for monomials of degree
four.
No new conditions appear.
(This implies that the ground-states for
chains of up to four sites can be obtained by this method.)
For monomials of
degree five, however, the consistency conditions give
that the traces of all monomials of degree two to four are zero.
%
%
Since we went up to monomials
of degree five one can guess what kind of
dirty algebra one had to do
(also \cite{Ho}).
We looked without success for conditions on the rates in order to
find non-zero
solutions.
The equations are however so cumbersome that we can't even be
sure that we didn't miss one.

We have also looked for finite-dimensional representations of the algebra
also with a negative result. As a by-product we have found that the
algebra (\ref{etii} exists if
%
\begin{equation}
g_{10}g_{01}=g_{20}g_{02}\,,\quad
g_{12}=-g_{21}=\frac{g_{20}(g_{02}-g_{01})}{g_{01}+g_{20}}
\end{equation}
Notice that this condition is incompatible with positivity of the rates.
This algebra has a two-dimensional representation:
%
\begin{eqnarray}
{\cal E}_0=\frac{1}{g_{20}-g_{01}}\left(\begin{array}{cc}
        g_{01}/g_{02}&0\\
        0&1
        \end{array}
        \right)\nonumber \\
{\cal E}_1=\frac{1}{g_{01}-g_{02}}\left(\begin{array}{cc}
        1&0\\
        -({g_{01}}^2+{g_{20}}^2)/ \lambda{g_{20}}^2&g_{01}/g_{20}
        \end{array}
        \right) \\
{\cal E}_2=\frac{1}{g_{02}-g_{01}}\left(\begin{array}{cc}
        g_{01}/g_{20}&\lambda\\
        0&1
        \end{array}
        \right)\nonumber \;.
\end{eqnarray}
Here $\lambda$ is an arbitrary parameter.

%
\subsection{$x_0=0$, $x_1$ and $x_2$ different from zero}
We define
\begin{equation}
\label{etvi}
D_1=x_1\, E_1\, ,\qquad D_2=x_2\, E_2
\end{equation}
and the algebra (\ref{eqvi}) becomes:
\begin{eqnarray}
g_{01}\, D_0\, E_1 - g_{10}\, E_1\, D_0 = D_0 \nonumber\\
g_{02}\, D_0\, E_2 - g_{20}\, E_2\, D_0 = D_0 \label{etvii}\\
g_{12}\, E_1\, E_2 - g_{21}\, E_2\, E_1 = E_1 - E_2 \nonumber
\end{eqnarray}
This algebra has a special structure in the sense that all the independent
monomials in $D_0$, $E_0$ and $E_2$
can be organized in the following way:
\begin{equation}
\label{etviii}
P_0\;,\;\;D_0 P_1\;,\;\;{D_0}^2 P_2\;, \ldots
\end{equation}
where the $P_i$ are monomials in $E_1$ and $E_2$ only.
This will imply in
the trace problem a decoupling of the recurrence relations according to the
power of $D_0$ appearing in the monomials.
In particular for words without
$D_0$'s, we can take
$D_0=0$ in Eq.(\ref{etvii}) and are left with the $N=2$ algebra
containing $E_1$ and $E_2$ for which we know that we have
one-dimensional
representations and thus in this sector the problem is solved.
The problem
is of course marrying the last equation in (\ref{etvii}) with the first two.
The
decoupling of the trace problem in various sectors will have also an
unexpected consequence in the representations of the algebra.
The
representations with a trace for words containing $D_0$'s will not have
a trace for words not containing any $D_0$.
So much for the structure of the
algebra (\ref{etvii}).

Going up to monomials of order three, the consistency relations obtained from
the equations giving
$\tr (D_0\, E_1\, E_2)$ and $\tr (E_2\, E_1\, D_0)$
give again Eq.(\ref{etiii})
(this is compatible with Eq.(\ref{eqx}) in which one can take $x_0=0$)
or
\begin{equation}
\label{etix}
g_{01}\, g_{02}= g_{10}\, g_{20}\;\; .
\end{equation}
We first assume
\begin{equation}
\label{etx}
g_{01}\, ,\;
g_{20}\, ,\;
g_{10}\, ,\; g_{02} \neq 0
\end{equation}
and look at words of order four.
One gets two new conditions which together
with (\ref{etix}) give:
\begin{equation}
\label{etxi}
g_{10}=g_{02}\, ,\qquad
g_{01}=g_{20}\, ,\qquad
g_{21}-g_{12}=g_{01}-g_{10}
\end{equation}

We will now introduce the following notations:
\begin{equation}
\label{etxii}
q=\frac{g_{01}}{g_{10}}=\frac{g_{20}}{g_{02}}\,,\qquad
r=\frac{g_{21}}{g_{12}}
\end{equation}
and
\begin{equation}
\label{etxiii}
E_1=\frac{G_1}{g_{01}-g_{10}}\, ;\; E_2=\frac{G_2}{g_{10}-g_{01}}\;\;.
\end{equation}
Taking into account the conditions (3.11) on the rates, the new algebra is:
\begin{eqnarray}
q\, D_0\, G_1 - G_1\, D_0 = (q-1)\, D_0            \nonumber\\
q\, G_2\, D_0 - D_0\, G_2 = (q-1)\, D_0            \label{etxiv}\\
r\, G_2 \,G_1 - G_1\, G_2 = (r-1)\, (G_1 + G_2)    \nonumber
\end{eqnarray}

At this point we are not going to look at words of order five or more but
show that a representation with a trace exits.
Before we show this, let us
first notice that the algebra (\ref{etxiv})
is invariant under the transformation:
\begin{equation}
\label{etxv}
D_0 \mapsto D_0 \, ,\; G_1 \mapsto G_2 \, ,\;
G_2 \mapsto G_1 \, ,\; q \mapsto \frac{1}{q} \, ,\; r \mapsto \frac{1}{r}
\end{equation}
and that one has the identities:
\begin{eqnarray}
q^n\, {D_0}^n \,G_1 - G_1\, {D_0}^n = (q^n -1)\, {D_0}^n    \nonumber\\
q^n\, G_2\, {D_0}^n - {D_0}^n\, G_2 = (q^n -1)\, {D_0}^n
\label{etxvi}
\end{eqnarray}
The Eqs.(\ref{etxv}) and (\ref{etxvi}) allow
to find traces on some monomials if one knows
some others.

In order to show that a representation exists, we first write
\begin{eqnarray}
G_1 = 1- \sqrt{r-1}\, {\cal A}         \nonumber\\
G_2 = 1+ \sqrt{r-1}\, {\cal B}         \label{etxvii}\\
D_0 = d_0 (1+(q-1) \, {\cal N})         \nonumber
\end{eqnarray}
where $d_0$ is an arbitrary parameter.
Using (\ref{etxiv}) we get:
\begin{eqnarray}
{\cal A}\, {\cal B} - r\, {\cal B} \,{\cal A} =1               \nonumber\\
{\cal A}\, {\cal N} - q\, {\cal N}\, {\cal A} ={\cal A}
\label{etxviii}\\
{\cal N}\, {\cal B} - q\, {\cal B} \,{\cal N} ={\cal B}        \nonumber
\end{eqnarray}
The algebra (\ref{etxviii}) which contains an $r$--deformed harmonic oscillator
(generators $\cal{A}$ and $\cal{B}$) together with $q$--deformed
actions of the number operator $\cal N$ has a Fock
representation \cite{AlDaRi,EsRi}:
\begin{equation}
\label{etxix}
{\cal A}\, |0> = <0|\, {\cal B}=0\; , \qquad {\cal B}={\cal A}^T
\end{equation}
where ${\cal A}^T$ is the transpose of ${\cal A}$,
\begin{equation}
\label{etxx}
{\cal A}=\left(\begin{array}{lllll}
        0&g_1&0&0&\cdots \\
        0&0&g_2&0& \\
        0&0&0&g_3 \\
        \vdots&&&\ddots&\ddots
        \end{array}
        \right)
\, ,\quad
{\cal N}=\left(\begin{array}{lllll}
        p_1&0&0&\cdots \\
        0&p_2&0& \\
        0&0&p_3 \\
        \vdots&&&\ddots
        \end{array}
        \right)
\end{equation}
and
\begin{equation}
\label{etxxi}
{g_n}^2=\{n\}_r,\qquad
p_n=\{n-1\}_q,\qquad
\{n\}_\lambda=\frac{\lambda^n-1}{\lambda -1}
\end{equation}
It is convenient to denote:
\begin{equation}
\label{etxxii}
G_1=1+{\cal F}_1,\qquad
G_2=1+{\cal F}_2,\qquad
D_0=d_0\, {\cal I}(q)
\end{equation}
The matrices ${\cal F}_1$, ${\cal F}_2$ and ${\cal I}(q)$ have a simple form:
\begin{equation}
\label{etxxiii}
\fl
{\cal F}_1=\left(\begin{array}{rrrrr}
        0&-f_1&0&0&\cdots \\
        0&0&-f_2&0& \\
        0&0&0&-f_3 \\
        \vdots&&&\ddots&\ddots
        \end{array}
        \right)
\, ,\quad
{\cal F}_2=-{\cal F}_1^T
\, ,\quad
{\cal I}(q)=\left(\begin{array}{lllll}
        e_1&0&0&\cdots \\
        0&e_2&0& \\
        0&0&e_3 \\
        \vdots&&&\ddots
        \end{array}
        \right)
\end{equation}
where
\begin{equation}
\label{etxxiv}
{f_k}^2=r^k-1,\qquad
e_k=q^{k-1}
\end{equation}
Let us now notice the following useful relations:
\begin{eqnarray}
q\, {\cal I}(q)\, {\cal F}_1 = {\cal F}_1\, {\cal I}(q) \nonumber\\
q\, {\cal F}_2\, {\cal I}(q) = {\cal I}(q)\, {\cal F}_2 \label{etxxv}
\end{eqnarray}
and
\begin{eqnarray}
{\cal F}_1\, {\cal F}_2 = 1 - r\, {\cal I}(r) \nonumber\\
{\cal F}_2 \,{\cal F}_1 = 1 - {\cal I}(r)    \label{etxxvi}
\end{eqnarray}
as well as
\begin{equation}
\label{etxxvii}
{\cal I}(r)\,{\cal I}(q)={\cal I}(rq)
\end{equation}
and
\begin{equation}
\label{etxxviii}
\tr\, {\cal I}(\lambda)=\frac1{1-\lambda}
\end{equation}

The calculation of the trace of any monomial containing at least one
$D_0$ proceeds as follows.
Using Eq.(\ref{etxxiii}) one has to compute only traces
containing an equal number of ${\cal F}_1$'s and
${\cal F}_2$'s together with ${\cal I}(q)$'s.
We use
the commutation relations (\ref{etxxv}) to bring the ${\cal I}(q)$'s
together and ``condense''
then into one using Eq.(\ref{etxxvii}).
Next we use Eq.(\ref{etxxvi}) in order to express
products of ${\cal F}_1$'s and
${\cal F}_2$'s in terms of ${\cal I}(r)$'s and make them ``condense''
together with the ${\cal I}(q^n)$ obtained from, let us say, $n$
${\cal I}(q)$'s.
The final result
is an expression which contains ${\cal I}$'s of various
arguments each one having a
trace given by Eq.(\ref{etxxviii}).
The $d_i$ of Eq.(\ref{eqai}) are $d_0$, $d_1=x_1$ and $d_2=x_2$.
This concludes our discussion of the algebra with
the conditions (\ref{etxi}).
The special case $q=r$ is already known and was
applied in Ref.\cite{DeJaLeSp}.

\vspace{0.5cm}
We now return to Eq.(\ref{etix}) and consider the case
\begin{equation}
\label{etxxix}
g_{01}=g_{20}=0
\end{equation}
We can find directly representations of the algebra in this case.

First we consider
\begin{equation}
\mu\equiv\frac{g_{10}}{g_{21}-g_{12}}\neq 1,\quad
\nu\equiv\frac{g_{02}}{g_{21}-g_{12}}\neq 1\;.
\end{equation}
We make a change of notations:
\begin{equation}
\label{etxxx}
E_1=-\frac{G_1}{g_{10}},\qquad
E_2=\frac{G_2}{g_{02}}
\end{equation}
and instead of the algebra (\ref{etvii}) we get:
\begin{eqnarray}
G_1\, D_0 = D_0 \nonumber\\
D_0\, G_2 = D_0 \label{etxxxi}\\
g_{12}\, G_1\, G_2 - g_{21}\, G_2\, G_1 = g_{02}\, G_1 + g_{10}\, G_2 \nonumber
\end{eqnarray}
Similar to what we did in the case of the last algebra (see Eq.(\ref{etxxii})),
we write:
\begin{equation}
\label{etxxxii}
G_1=1+{\cal F}_1,\qquad
G_2=1+{\cal F}_2,\qquad
D_0=d_0\, {\cal F}_0
\end{equation}
and make the observation that the Fock representation of the following
algebra \cite{AlDaRi,EsRi}:
\begin{eqnarray}
{\cal A}\, {\cal I}(0)= 0             \nonumber\\
{\cal I}(0)\, {\cal  B}= 0             \nonumber\\
\xi ({\cal A}\, {\cal  B} -r\, {\cal B} \,{\cal A})={\cal A}+{\cal B}+1
			         \label{etxxxiii}\\
{\cal A}\,|0>=<0|\,{\cal B}=0                  \nonumber\\
{\cal B}={\cal A}^T                            \nonumber
\end{eqnarray}
is known:
\begin{equation}
\label{etxxxiv}
{\cal A}=\left(\begin{array}{lllll}
	a_1&k_1&0&0&\cdots \\
	0&a_2&k_2&0& \\
	0&0&a_3&k_3 \\
	\vdots&&&\ddots&\ddots
	\end{array}
	\right)
\end{equation}
where
\begin{equation}
\label{etxxxv}
a_n=\frac1{\xi} \;\frac{r^{n-1}-1}{r-1},\qquad
{k_n}^2= \frac1{\xi^2}\; \frac{r^n-1}{r-1}
  	\left(\xi+\frac{r^{n-1}-1}{r-1} \right)\;\;,
\end{equation}
and ${\cal I}(q=0)$ as in (\ref{etxxiii}).
We introduce the following notations for ratios of rates
\begin{equation}
r=\frac{g_{21}}{g_{12}},\quad
\xi=\frac{\mu+\nu -1}{(\mu -1)(\nu -1)}\;\; .
\end{equation}
Using (\ref{etxxxiii})--(\ref{etxxxv}) we find a
representation for ${\cal F}_0$, ${\cal F}_1$ and
${\cal F}_2$
of Eq.(\ref{etxxxii}):
\begin{eqnarray}
{\cal F}_0={\cal I}(0)     \nonumber\\
{\cal F}_1=(\mu-1)({\cal I}(r)+{\cal V}-1) \label{etxxxvii}\\
{\cal F}_2=(\nu-1)({\cal I}(r)+{\cal V}^T-1) \nonumber
\end{eqnarray}
where ${\cal I}(r)$ is defined as above
and
\begin{equation}
\label{etxxxviii}
{\cal V}=\left(\begin{array}{lllll}
        0&s_1&0&0&\cdots \\
        0&0&s_2&0& \\
        0&0&0&s_3 \\
        \vdots&&&\ddots&\ddots
        \end{array}
        \right)
\end{equation}
with
\begin{equation}
\label{etxxxix}
{s_n}^2=(r^n-1)(\xi+r^{n-1}-1)\;\; .
\end{equation}
Notice that
\begin{equation}
r\; {\cal I}(r)\; {\cal V}= {\cal V}\; {\cal I}(r) \;\; ,
\end{equation}
and that the products ${\cal V}\,{\cal  V}^T$ and
${\cal V}^T\, {\cal V}$ can be written in terms of ${\cal I}(r)$
and ${\cal I}(r^2)$.
This makes the discussion of the existence and calculations of the
traces to be identical to that of the previous algebra.
A special case of
this algebra when $r=0$
was already discussed in Ref.\cite{Ma}.

For the case $\mu=1$ one must use a different representation
(the case $\nu$ is similar).
We write
\begin{equation}
\label{etxxxx}
\fl
E_1=\frac1{g_{10}}\left(1+\frac{g_{02}}{g_{02}-g_{10}}{\cal A}\right)\qquad
E_2=-\frac1{g_{02}}\left(1+\frac{g_{02}-g_{10}}{g_{12}}{\cal B}\right)\qquad
D_0=d_0\, {\cal I}(0)
\end{equation}
where now ${\cal A}$ and ${\cal B}$ fulfill:
\begin{eqnarray}
{\cal A}\, {\cal I}(0)= 0             \nonumber\\
{\cal I}(0)\, {\cal  B}= 0             \nonumber\\
{\cal A}\, {\cal  B} -r\, {\cal B} \,{\cal A}={\cal A}+1
			         \\
{\cal A}\,|0>=<0|\,{\cal B}=0
\end{eqnarray}
The matrices are given by
\begin{equation}
{\cal A}=\left(\begin{array}{lllll}
	0&f_1&0&0&\cdots \\
	0&0&f_2&0& \\
	0&0&0&f_3 \\
	0&0&0&0&\ddots \\
	\vdots&&&&\ddots
	\end{array}
	\right),
\quad
{\cal B}=\left(\begin{array}{lllll}
        b_1&0&0&0&\cdots \\
        f_1&b_2&0&0& \\
        0&f_2&b_3&0 \\
        0&0&f_3&b_4 \\
        \vdots&&&\ddots&\ddots
        \end{array}
        \right)
\end{equation}
with
\begin{equation}
b_n=\frac{r^{n-1}-1}{r-1},\qquad
{f_n}^2= \frac{r^n-1}{r-1}
\end{equation}
Compare with Eqs.(\ref{etxxxiii})--(\ref{etxxxv}).

The profound reason which explains why the same representations
for the $E_\alpha$ occur in the problem with periodic boundary
conditions and for the open end problems (see Ref.\cite{AlDaRi}, Sec.7)
comes from the following observation.
The differences between the two types of boundary conditions is that
the parameters $x_1$ and $x_2$ are free for the periodic case
but are determined by the boundary matrices in the open case.
In the latter case supplementary relations exist among the bulk
($g_{\alpha\beta}$) and boundary rates. Since the $E_{\alpha}$
found in the case of open ends have a trace, they can be used in the
present paper.

\subsection{$x_0=x_1=0$, $x_2$ different from zero}
In this case the algebra (\ref{eqvi}) becomes:
\begin{eqnarray}
g_{02}\,D_0\,E_2-g_{20}\,E_2\,D_0=D_0 \nonumber\\
g_{12}\,D_1\,E_2-g_{21}\,E_2\,D_1=D_1 \label{etxxxxi}\\
g_{01}\,D_0\,D_1-g_{10}\,D_1\,D_0=0   \nonumber
\end{eqnarray}
where we have taken $D_2=x_2\,E_2$.
If one wants non-trivial ground states with particles of the
three species, one
has to take:
\begin{equation}
\label{etxxxxii}
g_{01}=g_{10}=0
\end{equation}
In this case the symmetry of the problem is enormous since no prescription
is given on products of $D_0$'s and $D_1$'s.
The right language are affine
symmetries \cite{Ar} and it is beyond the scope of this paper to get involved
in the representation theory for this case.
The algebra with traces however
very probably exists since it was used already in Ref. \cite{Ev}
in the sector of one
$D_0$ and one $D_1$ and there are no obvious reasons why it should not be so
in the general case.
This model was also investigated in \cite{MeBaDh} by different methods.

Finally the case $x_0=x_1=x_2=0$ takes us the situation of symmetric rates
where, as we already know, we get a symmetric wave function.

\section{Conclusion}
Leaving aside physical applications, our own fascination with the problem
described in this paper comes from
the unusual properties of the representations
of the algebra appearing in searching for matrix-products ground-states.
As stressed already in the introduction certain monomials or even all
monomials of the same degree in $D$'s
have, up to a normalisation factor, traces
independent of the representation.
More has to be understood in the
general case when the $X$'s are matrices (see Eq.(\ref{eqiii}))
and also in the class
of algebras given by Eq.(\ref{eqvi}).
It is probably possible to encode the
conditions on the rates $g_{\alpha\beta}$ in some cubic and quartic
identities.
This guess is based on the fact that conditions found on the
$g_{\alpha\beta}$ were
obtained from words of degree three and four.
For this reason the
four-state problem is worth looking at in order to see if a general
pattern appears.
{}From the point of view of physical applications, for the three-state
problem we have solved the recurrence relations if the rates satisfy the
conditions (\ref{etiii}), (\ref{etxi}) or (\ref{etxxix}).
The case given by Eq.(\ref{etxxxxii}) was known
already.
No other cases exist.
However, as everybody knows from solving
the recurrence relations to computing a relevant physical
quantity it is a long way to go.

\ack
We would like to thank SISSA and our colleagues here for the warm
and stimulating environment
and the EU and DAAD for
financial support.
We are grateful to S.~Dasmahapatra,
B.~Derrida, B.~Dubrovin, A.~Honecker, K.~Mallick and
P.~Martin for discussions.

%

%
\end{document}